\documentclass[showpacs,floatfix,superscriptaddress,twocolumn,prb]{revtex4}
\usepackage{amsfonts}
\usepackage{stmaryrd}
\usepackage{bbm}
\usepackage{mathrsfs}
\usepackage{tipa}
\usepackage{amssymb}
\usepackage{txfonts}

\usepackage{graphicx}
\usepackage{dcolumn}
\begin{document}

\newcommand*{\cm}{cm$^{-1}$\,}

\title{Coexistence and competition of multiple charge-density-wave orders in rare-earth tri-telluride RTe$_3$}

\author{B. F. Hu}
\affiliation{Key Laboratory of Neutron Physics, Institute of Nuclear
Physics and Chemistry, China Academy of Engineering Physics,
Mianyang 621900, China}
\affiliation{Beijing National Laboratory for
Condensed Matter Physics, Institute of Physics, Chinese Academy of
Sciences, Beijing 100190, China}

\author{B. Cheng}
\author{R. H. Yuan}
\author{T. Dong}
\affiliation{Beijing National Laboratory for Condensed Matter
Physics, Institute of Physics, Chinese Academy of Sciences, Beijing
100190, China}

\author{N. L. Wang}

\affiliation{Beijing National
Laboratory for Condensed Matter Physics, Institute of Physics,
Chinese Academy of Sciences, Beijing 100190, China}
\affiliation{Collaborative Innovation Center of Quantum Matter, Beijing, China}
\affiliation{International Center for Quantum Materials, School of Physics, Peking University, Beijing 100871, China}

\begin{abstract}
The occurrences of collective quantum states, such as
superconductivity (SC) and charge- or spin-density-waves (CDWs or
SDWs), are among the most fascinating phenomena in solids. To date
much effort has been made to explore the interplay between different
orders, yet little is known about the relationship of multiple
orders of the same type. Here we report optical spectroscopy study
on CDWs in the rare-earth tri-telluride compounds \emph{R}Te$_3$
(\emph{R} = rare earth elements). Besides the prior reported two CDW
orders, the study reveals unexpectedly the presence of a third CDW
order in the series which evolves systematically with the size of
\emph{R} element. With increased chemical pressure, the first and
third CDW orders are both substantially suppressed and compete with
the second one by depleting the low energy spectral weight. A
complete phase diagram for the multiple CDW orders in this series is
established.
\end{abstract}


\maketitle

\section{Introduction}
Charge-density-wave (CDW) states in low-dimensional electronic
systems are among the most actively studied phenomena in condensed
matter physics. Most CDW states are driven by the nesting topology
of Fermi surfaces (FSs), i.e., the matching of sections of FS to
others by a wave vector \emph{\textbf{q}} = 2\emph{\textbf{k}$_F$},
where the electronic susceptibility has a divergence.
\cite{dynamics,densitywave} A single particle energy gap opens in
the nested regions of the FSs at the transition, which leads to the
lowering of the electronic energies of the system. Coupled to the
lattice by electron-phonon interactions, the development of CDW
state also causes a lattice distortion with the superstructure
modulation wave vector matching with the FS nesting wave vector.

The family of rare-earth tri-telluride \emph{R}Te$_3$ (\emph{R}
being an element of the lanthanide family) presents an excellent
low-dimensional model system to study the effect of FS
nesting-driven CDW formation. \emph{R}Te$_3$ has a layered structure
consisting of alternate stacking of the insulating \emph{R}Te slab
and the conducting Te only double planes along \emph{b} axis.
\cite{structure1,structure2} The FSs are strongly two-dimensional
(2D) cylindrical like and exhibit nesting instabilities, leading to
CDW ground states. \cite{ARPESSmTe3,theoryRTe31,R-evolution3}
Interestingly, the CDW properties can be well tuned by choosing
different elements of the lanthanide series. As the lanthanide
4\emph{f} electrons are far below the Fermi level, the major effect
of changing different lanthanide element, without the entanglement
of charge doping effect, is to exert chemical pressure.
\cite{R-evolution1,R-evolution2,2CDW} By moving through the series
from La to Tm, the increasing occupation of 4\emph{f} orbital leads
to a decrease of the ion radii and the lattice parameters.
\cite{R-evolution2} For \emph{R}Te$_3$, an incommensurate CDW ground
state with a wave vector \emph{$\textbf{q}_1$} $\thickapprox$ 2/7
\emph{$\textbf{c}^{\ast}$} was commonly observed.
\cite{R-evolution1,R-evolution2} For the four heavy rare-earth
\emph{R}Te$_3$ (\emph{R}=Tm, Er, Ho, Dy) compounds, the development of a second CDW order,
with the wave vector \emph{$\textbf{q}_2$} $\thickapprox$ 1/3
\emph{$\textbf{a}^{\ast}$} perpendicular to the first one, was
revealed and well documented. \cite{2CDW,ARPES2CDW,opticalErTe3}

In our previous optical spectroscopy study on CeTe$_3$ and TbTe$_3$,
we also observed a clear, though weak, CDW energy gap feature
developing below roughly 200 K, besides the major energy gap
structure at higher energy, \cite{opticalCeTe3,opticalTbTe3}
suggesting presence of another CDW order even in the light and
intermediate rare-earth tri-telluride compounds. However, the gap
amplitude does not follow the trend observed for the four heavy
rare-earth \emph{R}Te$_3$ compounds. \cite{R-evolution3} Those
findings were extremely puzzling and motivated us to conduct further
systematic study. In this work we report the in-plane optical study
on all of the eleven \emph{R}Te$_3$ (\emph{R} = La - Nd, Sm, Gd -
Tm) compounds. The measurement clearly reveals the coexistence of
multiple CDW orders in all members of \emph{R}Te$_3$ family. Besides
the prior reported two ones, our optical study unexpectedly
discovers the presence of a third CDW order in the series. The
energy gaps observed previously in CeTe$_3$ and TbTe$_3$ at lower
energies actually belong to the third CDW order which had never been
reported by any other probes before. The first and third CDW orders
exhibit much similar systematic evolution as a function of \emph{R}
and cooperate with each other while both compete with the second CDW
order. We suggest that the third CDW order arises from the bilayer
splitting, which lifts the degeneracy of conduction bands of double
Te sheets. A complete phase diagram of CDW energy gaps versus
lanthanide elements was established for the \emph{R}Te$_3$
compounds.

\section{\label{sec:level2}EXPERIMENT AND RESULTS}
The single crystals in the present study were grown by a self-flux
method with a molar ratio \emph{R}:Te = 1:40 in a procedure the same
as in reference 13. The optical reflectance measurements were
conducted on the Bruker IFS 80 v/s spectrometer in a frequency range
from 40 \cm to 25,000 \cm. An \textit{in situ} gold and aluminium
overcoating technique was used to get the reflectivity
\emph{R($\omega$)}. \cite{Homes} The real part of the conductivity
spectra $\sigma_1(\omega)$ was obtained through the Kramers-Kronig
transformation of \emph{R($\omega$)}. A Hagen-Rubens relation was
used in the low frequency extrapolation and in the high frequency
part a constant value extrapolation was used up to 100,000 \cm,
above which an $\omega^{-4}$ relation was employed.

\begin{figure*}[t]
{\includegraphics[width=18cm,clip]{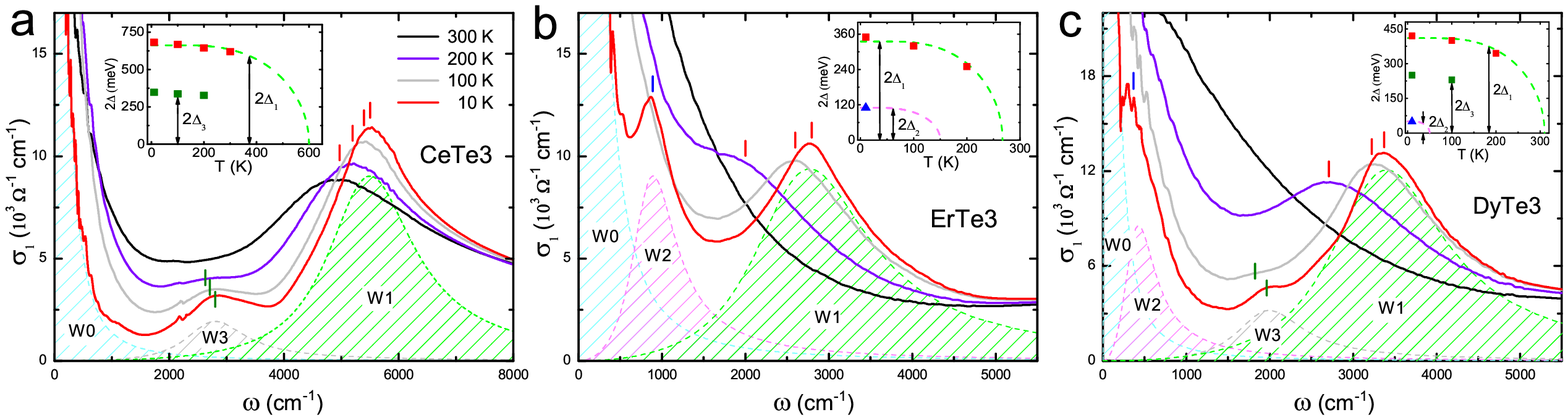}} \caption{(Color
online) Temperature dependent optical conductivity
$\sigma_1(\omega)$ of CeTe$_3$ (a), ErTe$_3$ (b) and DyTe$_3$ (c).
The fitting curve of each Lorentz oscillation mode at 10 K, as well
as that of the Drude resonance, was plotted at the bottom. The
corresponding spectral weight, \emph{W0} (Drude), \emph{W1} (gap 1),
\emph{W2} (gap 2) and \emph{W3} (gap 3), were revealed by the dashed
area. The gap sizes were indicated by the short vertical lines.
Insets: The CDW gaps' sizes as functions of T. The dashed lines show
the gap function 2\emph{$\Delta$(T)} based on the weak coupling mean
field theory and serve as a guide to the eye. The values
2\emph{$\Delta$(T)} are scaled to the experimental results and
adjusted so as to fit the gap sizes at different temperatures.}
\end{figure*}

We have performed temperature dependent optical measurements on all
of the eleven compounds in rare-earth tri-telluride family
\emph{R}Te$_3$ (\emph{R} = La - Nd, Sm, Gd - Tm). Figure 1 shows
temperature dependent optical spectra of three representative
compounds CeTe$_3$, ErTe$_3$ and DyTe$_3$. We note that the
development of multiple peak features at low temperature not only
appears in the heavy rare-earth compounds but also emerges in the
light ones. In DyTe$_3$ even three peaks are present in the conductivity spectrum
at 10 K. Meanwhile, the residual Drude component narrows significantly.
The data yield explicit evidence for the opening of multiple partial
CDW gaps on FSs.\cite{opticalErTe3,opticalCeTe3} Usually,
the peak position was used to estimate the CDW energy gap due to the
spectral weight transfer from the free carrier
response to the energy scale just above the energy gap of $\hbar\omega$=2$\Delta$.
The formation of the peak or maximum in $\sigma_1(\omega)$ is caused by
both the density of state and the type I coherence factor effect.\cite{dynamics,densitywave}
To quantify the discussions, a Drude-Lorentz model was employed to
extract the CDW gap sizes and the relevant spectral weight,
\cite{opticalErTe3,opticalCeTe3,opticalTbTe3} as will be presented below in detail.

Figure 2 shows the optical spectra of the whole \emph{R}Te$_3$
series at two representative temperatures 10 K and 300 K. For each
compound multiple suppressions in \emph{R($\omega$)} arise at 10 K
and simultaneously multiple peaks appear in $\sigma_1(\omega)$. Both
features suggest the formations of CDW orders. The CDW energy gaps
were indicated by the short vertical lines and could be obviously
categorized into three groups: gap 1, gap 2 and gap 3. In each
group, the gap amplitude exhibits monotonic evolution as a function
of chemical pressure (\emph{R} element). The Drude-Lorentz fitting
results, as well as the CDW transition temperatures, were collected
in Table 1 and a direct view of the systematic evolutions as
functions of \emph{R} was given in Fig. 3. At the lowest measurement
temperature gap 1 exists in the whole \emph{R}Te$_3$ series,
gap 2 only arises in the four heavy rare-earth compounds, while gap 3
survives from the light to relatively heavy ones. Specially in DyTe$_3$, all the three
gaps coexist. The first and second CDW orders occur with transition
temperatures \emph{T$_{c1}$} = 310 K and \emph{T$_{c2}$} = 52 K
respectively. \cite{2CDW} In our optical measurements the feature of
gap 3 arises between 100 K and 200 K while the transition
temperature has not yet been determined by other probes. The three CDW orders seem to
coexist also in the neighboring compound TbTe$_3$, as a recent synchrotron x-ray diffraction
study on TbTe$_3$ indicated the presence of the second CDW order with
transition temperature \emph{T$_{c2}$} = 41 K \cite{TbTe3}.
According to gap 2's evolution as a function of \emph{R},
the gap energy scale should be less than 50 meV. It
is rather close to the sizable Drude component and possesses much
smaller spectral weight. As a consequence, the gap structure may become blurred in
our optical measurement.

\begin{figure*}[t]
{\includegraphics[width=17cm,clip]{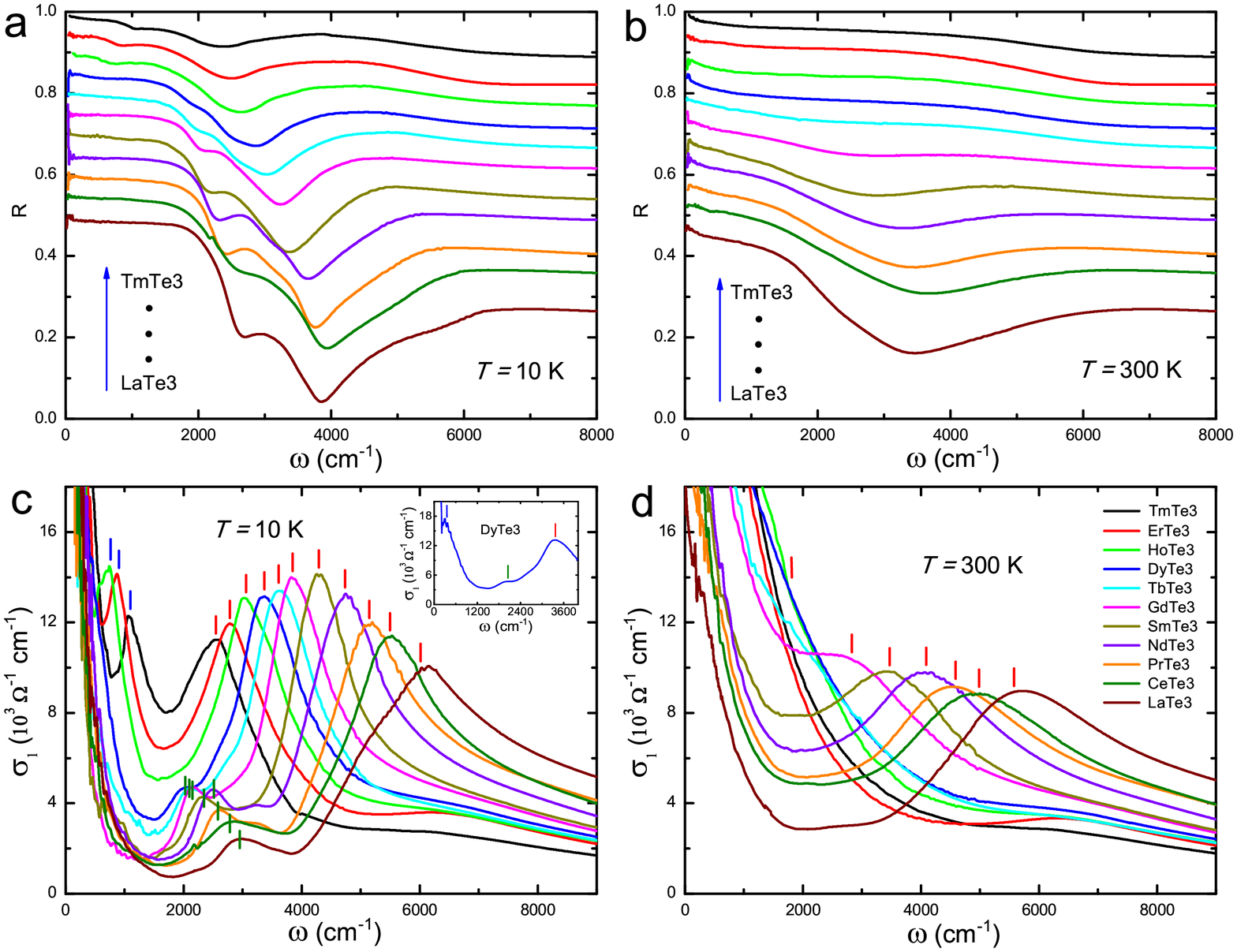}} \caption{(Color
online) (a) The in-plane optical reflectivity \emph{R($\omega$)} of
\emph{R}Te$_3$ at 10 K. (b) The same plot as (a) at 300 K. In both
panels, each \emph{R($\omega$)} curve, except for that of TmTe$_3$,
was shifted away vertically from its neighbors by 0.05 for clarity.
(c) The optical conductivity $\sigma_1(\omega)$ of \emph{R}Te$_3$ at
10 K. For each compound in \emph{R}Te$_3$ series, it clearly
indicates formations of multiple CDW energy gaps at 10 K, which
manifest strongly systematic evolutions across the series. According
to the peak position and evolution behavior, the multiple energy
gaps could be classified into three groups. The gap sizes were
indicated by the short vertical lines in red (gap 1), blue (gap 2)
and olive (gap 3). The inset shows the three gap features in
DyTe$_3$. (d) The optical conductivity $\sigma_1(\omega)$ of
\emph{R}Te$_3$ at 300 K. For the light rare-earth compounds from
LaTe$_3$ to TbTe$_3$, the first energy gap feature is still present
at 300 K.}
\end{figure*}

On traversing the lanthanide series from light rare-earth to heavier
ones, the lattice parameter \emph{a} decreases monotonically and
thus chemical pressure oppositely increases.
\cite{R-evolution2,2CDW} It is noteworthy that the \emph{R}Te$_3$
compounds are ideal platforms for the study of (chemical) pressure
tuned variations since doping entanglement is completely absent. \cite{2CDW,
quenchCDW} The 4\emph{f} electrons hide in the inner-shell and the
relative bands are far below the Fermi level, \cite{R-evolution3}
which have little influence over the FS properties. In Fig. 3(a) we
note that gap 1 and gap 3 show much similar monotonic evolutions and
both suffer substantial suppressions with increased chemical
pressure. For the heavy rare-earth compounds HoTe$_3$, ErTe$_3$ and
TmTe$_3$, gap 3 is completely suppressed. Meanwhile, gap 2 suddenly
arises and becomes larger against the other two ones from DyTe$_3$
to TmTe$_3$. The transition temperatures \emph{T$_{c1}$} and
\emph{T$_{c2}$} hold nearly the same evolution trend with the CDW
gaps. \cite{2CDW} In the spectral weight plot (Fig. 3(d)), \emph{W1}
and \emph{W3} manifest little variations versus \emph{a} for the
light rare-earth compounds. With the emergence of gap 2, both
undergo sudden depressions. The CDW versus \emph{a} plot establishes
a complete electronic phase diagram in \emph{R}Te$_3$ series, which
clearly reveals the coexistence and competition of multiple CDW
orders.

\begin{table*}[htbp]
\begin{center}
\newsavebox{\tablebox}
\begin{lrbox}{\tablebox}
\begin{tabular}{p{15mm}<{\centering}*{17}{p{10mm}<{\centering}}}
\hline \hline\\[1pt]
{}&LaTe$_3$&CeTe$_3$&PrTe$_3$&NdTe$_3$&SmTe$_3$&GdTe$_3$&TbTe$_3$&DyTe$_3$&HoTe$_3$&ErTe$_3$&TmTe$_3$\\[4pt]
\hline\\[4pt]
$2\Delta_1$(10 K)&750&680&640&590&530&480&450&420&380&350&320\\[4pt]
$2\Delta_2$(10 K)&--&--&--&--&--&--&--&50&90&110&140\\[4pt]
$2\Delta_3$(10 K)&370&350&320&310&290&270&260&250&--&--&--\\[4pt]
$2\Delta_1$(300 K)&700&620&570&510&430&350&220&--&--&--&--\\\\[4pt]
\emph{T$_{c1}$}&--&--&--&--&416&377&336&310&288&267&244\\[4pt]
\emph{T$_{c2}$}&--&--&--&--&--&--&--&52&110&157&180\\[4pt]
\hline \hline
\end{tabular}

\end{lrbox}
\caption{Single particle gap 2\emph{$\Delta$} and transition
temperature \emph{T$_c$} of the CDW orders in \emph{R}Te$_3$
compounds. The values in the top three rows and in the fourth one
are the CDW single particle gaps at 10 K and 300 K respectively. The
transition temperatures were defined by the transport anomaly in
\emph{$\rho(T)$} in reference 10. The transition temperatures
\emph{T$_{c3}$}s have not yet been reported.}
\scalebox{1.1}{\usebox{\tablebox}}
\end{center}
\end{table*}

\begin{figure*}
\includegraphics[width=17cm,clip]{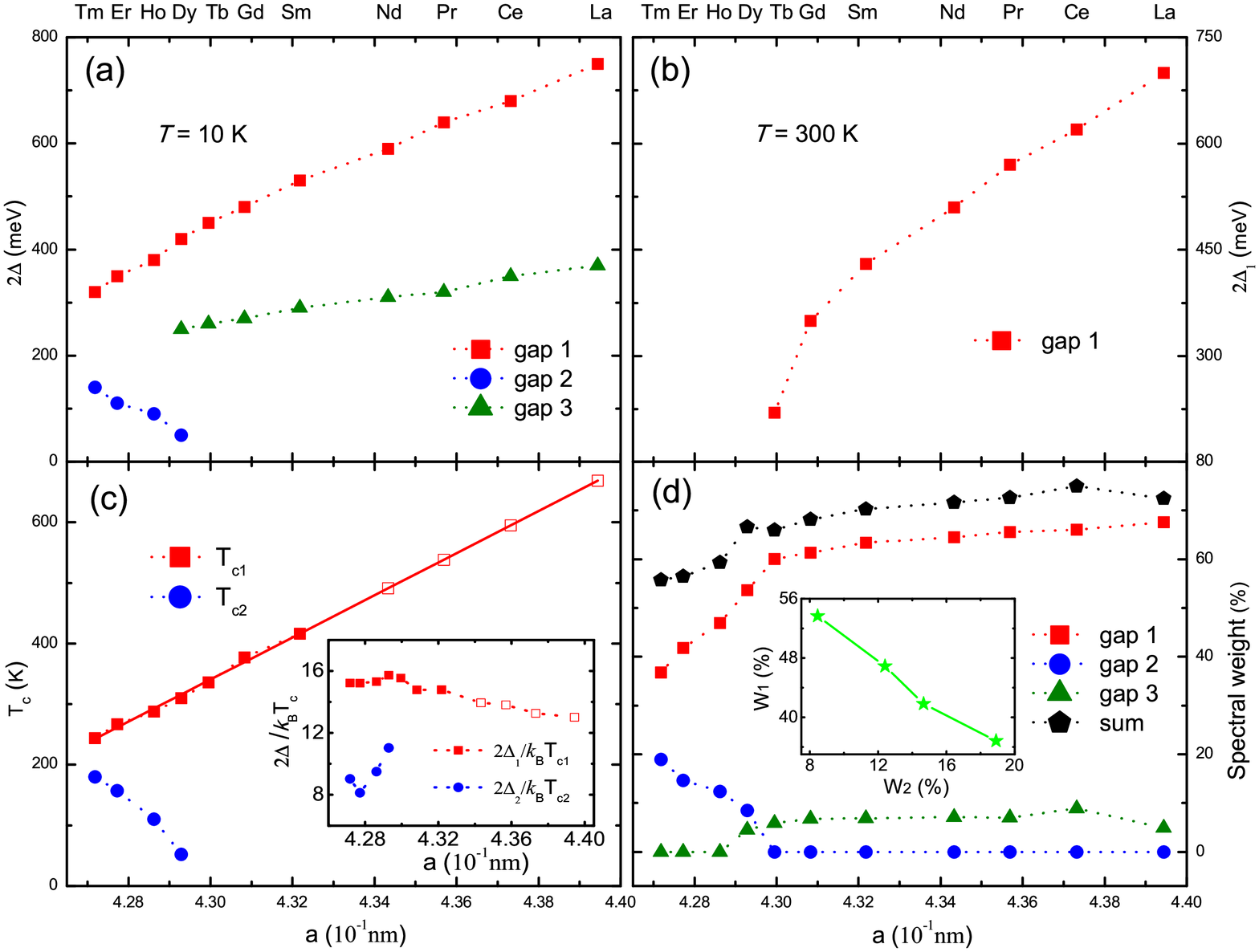}
\caption{(Color online) (a) and (b) CDW single particle gap
2\emph{$\Delta$} at 10 K and 300 K, respectively. (c) The transition
temperatures \emph{T$_{c1}$} and \emph{T$_{c2}$}. \cite{2CDW} For
the four light rare-earth compounds from LaTe$_3$ to NdTe$_3$,
\emph{T$_{c1}$}s are above 450 K and have not yet been determined by
experimental probes. The four values, which were plotted in open
squares, were obtained by the fitting and extrapolation of the other
compounds' transitions. The solid straight line is an indication of
the linear fit. The inset shows the ratios 2\emph{$\Delta/k_BT_c$}s
for the first and second CDW orders. (d) Chemical pressure
(\emph{R}) dependence of the spectral weight lost to each CDW gap
(\emph{W1}, \emph{W2}, \emph{W3}), expressed as a percentage of that
of the total free carriers in normal state. The total spectral
weight lost in CDW states is plotted in black pentagons. The inset
shows the plot of \emph{W$_1$} against \emph{W$_2$}, which clearly
demonstrates the anti-correlation of the two quantities. The lattice
parameter \emph{a} was obtained in reference 9 at 300 K, for which
the relative \emph{R}Te$_3$ compounds were indicated on the top of
the figures. The dotted lines were employed as a guide to the eye.}
\end{figure*}

Gap 1 is relative to the first CDW order,
\cite{opticalCeTe3,R-evolution3,ARPESSmTe3,ARPESCeTe3} which occurs
with an incommensurate wave vector \emph{$\textbf{q}_1$}
$\thickapprox$ 2/7 \emph{$\textbf{c}^{\ast}$}.
\cite{R-evolution1,R-evolution2} The transition temperature
\emph{T$_{c1}$} was plotted in Fig. 3(c). The second CDW order with
a wave vector \emph{$\textbf{q}_2$} $\thickapprox$ 1/3
\emph{$\textbf{a}^{\ast}$}, which is also incommensurate, only
arises in the four heavy rare-earth compounds from DyTe$_3$ to
TmTe$_3$ and is responsible for gap 2.
\cite{2CDW,ARPES2CDW,opticalErTe3} The FS nesting conditions
responsible for the two orders were illustrated in Fig. 4.
\cite{2CDW,theoryRTe31,theoryRTe32} By contrast, except for our
earlier optical probes on CeTe$_3$ and TbTe$_3$,
\cite{opticalCeTe3,opticalTbTe3} remarkably nothing about gap 3 was
known to date. Since the gap amplitude and its evolution in the
systems have been well established, the gap origin and its position
in \textbf{\emph{k}}-space are highly desired. Here, we would like to stress that
gap 3 does not belong to the other two known CDW orders. In the
spin-density-wave (SDW) transition in Fe-based superconducting
parents, two distinct energy scales were identified below
\emph{T$_c$}, \cite{opticalBaFe2As2} which were explained to arise
from the gapping of different FS sheets. Here the possibility was
ruled out because the three kinds of energy gaps all develop at
different temperatures. In our temperature dependent optical
measurements, the features relative to gap 3 all emerge below 300 K.
Moreover, gap 3, appearing between 2,000 \cm to 3,000 \cm, can not
be ascribed to the CDW collective excitations, either the phase mode
or the amplitude mode. The former is usually pinned in the microwave
or millimeter-wave spectral range by impurity or lattice
imperfections, \cite{dynamics,densitywave,electrodynamics} while the
latter is much lower than the unrenormalized phonon frequency at
Kohn anomaly. \cite{densitywave,phononRTe3} Furthermore, the
scenario of pseudogap character prior to the underlying second CDW
transition , due to fluctuation effect,
\cite{2CDW,opticalErTe3,densitywave} could also be excluded since
gap 2 and gap 3 exhibit the opposite evolution across the
lanthanides series and both features are present in DyTe$_3$.
Therefore, gap 3 represents a new or third CDW order, of which the
nesting condition is expected.

\begin{figure}
\includegraphics[width=9cm,clip]{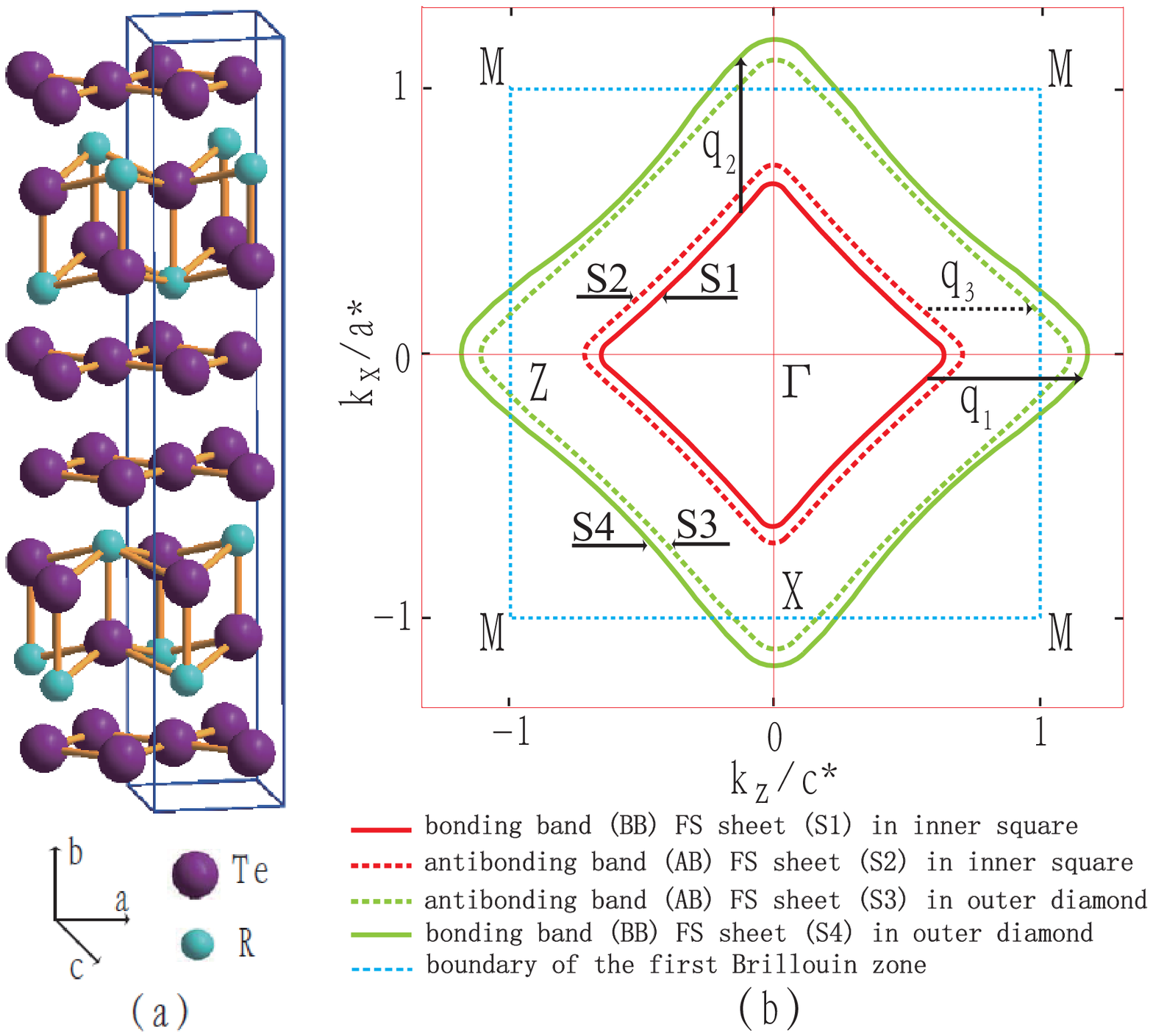}
\caption{(Color online) (a) Crystal structure of \emph{R}Te$_3$. (b)
Schematic diagram of FSs of \emph{R}Te$_3$. The FSs, showing little
dispersion along \emph{\textbf{b}$^{\ast}$} axis, are projected onto
the $\Gamma$-X-Z plane (\textbf{\emph{k$_y$}} = 0). The FS contours
consist of two $\Gamma$-centered pockets, the hole-like inner square
(in red) and the electron-like outer diamond (in green).
\cite{2CDW,theoryRTe31} Due to coupling of two neighboring Te
layers, each contour splits into two parallel sections (bilayer
splitting). The amplitude of bilayer splitting varies on the FS,
which is schematically illustrated by the separated distance. The
bilayer splitting reaches maximum near the corner ($\delta \approx
0.03 \textbf{c}^{\ast}$). \cite{R-evolution3} The nesting wave
vectors are also illustrated. For clarity, only $\Gamma$-centered
FSs are displayed.}
\end{figure}

The crystal structure \cite{structure1,structure2} and schematic FS
of \emph{R}Te$_3$ are displayed in Fig. 4.
\cite{2CDW,ARPES2CDW,R-evolution3,theoryRTe31,theoryRTe32} The plot of the
FSs has been simplified for the purpose of illustrating the nesting conditions.
According to the band structure calculation on the 3D
crystal structure with the employ of the linear muffin-tin orbital (LMTO) method, the
FS consists of three parts contributed by the p$_x$ and p$_z$ orbitals of metallic Te
layers: $\Gamma$-centered square like FS, M-centered FS mainly in the
second Brillouin zone and X/Z-centered small-size FS. Here we have unfolded the third part into the second Brillouin zone.
Then, the FSs could be approximately viewed as two $\Gamma$-centered
pieces: inner and outer ones.\cite{theoryRTe31,R-evolution3} The
4\emph{f} electrons of rare earth R are far below the Fermi level and have little
effect on \emph{E$_F$}. Since there are two conducting Te planes between
the insulating \emph{R}Te slabs in the crystal structure (as can be seen from Fig. 4(a)),
the interlayer coupling of the two neighboring metallic Te layers would break
the degeneracy of conduction band and yields two nearly parallel FS
crossings: bonding band (BB) and antibonding band (AB), which is
usually called bilayer splitting.
\cite{2CDW,R-evolution3,theoryRTe31,theoryRTe32} Then the FSs are
all double-wall like. It is noted that the outer FSs come from the band folding
effect due to the fact that the real three-dimensional (3D) lattice
structure actually doubles the unit cell of Te square lattice of the
Te layers along its diagonal direction, \cite{R-evolution3} then the
bilayer splitting of inner FSs (S1 and S2) and the outer FSs (S3 and
S4) are just opposite. Since S1 and S2 are nearly parallel and have
quite similar FS topology, and the same case applies to S3 and S4,
the nesting between the inner square and outer diamond, which is
along \emph{\textbf{c}$^{\ast}$} axis, would have two possible
selections: (a) S1 nests S3 and S2 nests S4, (b) S1 nests S4 and S2
nests S3. In case (a), the nesting takes place between two different
bands (interband nesting between BB and AB) and leads to a single
nesting wave vector, and thus a single CDW transition. It is
reminiscent of the blue bronze which, owning two partially filled
bands and thus four FS sheets, has only one CDW and undergoes a
metal-semiconductor transition. \cite{KMO,chemical-review} In case
(b), the nesting happens between the two FS sheets within each band
(intraband nesting within BB and AB), which implies two different
nesting wave vectors and two separated CDW transitions, as well as
two distinct CDW gap energy scales. Our study favors the latter
case.

Angle-resolved photoemission spectroscopy (ARPES) measurements
reveal that the FSs on both BB and AB bands, which locate near the
corner along \emph{\textbf{c}$^{\ast}$} axis, were completely
removed at 10 K. \cite{ARPES2CDW,R-evolution3,ARPESCeTe3} Meanwhile,
two different energy gaps were observed on the two parallel band
sheets, a larger gap on BB and the smaller on AB.
\cite{R-evolution3} Therefore, it could be concluded that the
nesting within BB leads to the first CDW order and that in AB causes
the third one. The two CDW orders both occur along
\emph{\textbf{c}$^{\ast}$} axis and the nesting vectors
\emph{\textbf{q}$_1$} and \emph{\textbf{q}$_3$} differ very little
since the bilayer splitting is rather small. The interpretation is
supported by the recent studies of the single layer rare-earth
di-telluride CeTe$_2$, \cite{CeTe2} where only one CDW order was
identified along \emph{\textbf{c}$^{\ast}$} axis since the bilayer
splitting is absent. \cite{theoryRTe31} It is worthy noting that Yao
\emph{et al}. have addressed the question of occurrence of just one
ordering transition or two distinct ones, which was determined by
the comparison of relevant energy scales \emph{t$_{bil}$}
(responsible for the bilayer splitting) and the CDW gap 2$\Delta$.
\cite{theoryRTe32} The present work clearly indicates that in
\emph{R}Te$_3$ series bilayer splitting is of primary importance and it
can result in separated CDW orders.

\section{\label{sec:level2}DISCUSSION}
Our optical study explicitly reveals the development of a third
energy gap in \emph{R}Te$_3$ series upon lowering T, which yields
strong evidence for the existence of a new CDW order distinct from
the prior two ones. Surprisingly, this order has never been
identified by any other techniques before. In the magnetic
susceptibility and transport probes, neither measurement observes
clear anomaly in consequence of formation of the third CDW order.
\cite{2CDW,transport,magnetism1,magnetism2} In Fig. 1, we notice
that, in contrast to the pronounced free carrier response and the
sizable spectral weight of the other two orders, the peak signature
ascribed to gap 3 is much less notable. In Fig. 3(d) we find that
just a very small amount of electrons response the third CDW
transition. Take CeTe$_3$ as an example. The spectral weight of
total free carriers in normal state is \emph{W = W0 + W1 + W2 + W3}
and the percentage of the spectral weight lost to each CDW gap is
\emph{$\Phi_i$ = W$_i$/W}. The fitting results show that
\emph{$\Phi_1$} $\approx$ 66\% while \emph{$\Phi_3$} $\approx$
8.9\%, which means that about two thirds FSs were gapped away in the
first CDW transition while in the third one the lost density of
states (DOS) at \emph{E$_F$} is substantially reduced. The very
small spectral weight of gap 3 gives us some clues to understand the
discrepancy. In magnetic susceptibility measurement of LaTe$_3$,
$\chi$ exhibits constant diamagnetic behavior.
\cite{transport} La and Te are both heavy atoms with many closed
shell core electrons. They collectively contribute considerable
diamagnetism which overcomes the small Pauli paramagnetism of the
free electron gas. For the other compounds in \emph{R}Te$_3$, $\chi$
is dominated by the 4\emph{f} electrons and exhibits Curie-Weiss
paramagnetism. \cite{magnetism1,magnetism2} The effect of little
variation of DOS near \emph{E$_F$} due to the third CDW is
significantly reduced. The notion could also apply to the transport
measurement, where resistivity \emph{$\rho$(T)} is
determined by several factors, e.g. Fermi velocity, scattering rate and free carrier
density. The reduction of DOS near \emph{E$_F$} might be compensated by
the other two factors. As a consequence, this CDW order could have rather weak effect
on \emph{$\rho$(T)} and the corresponding signature becomes
obscured.

Since the bilayer splitting is very small, the two wave vectors
\emph{q$_1$} and \emph{q$_3$}, as well as their associated
modulation periodicity, are very close to each other. Moreover, the
scattering intensity of CDW superlattice peaks is several orders of
magnitude smaller than that of the average structure Bragg peaks.
\cite{2CDW} To distinguish between the first and third CDW orders
becomes extremely hard in X-ray scattering measurement. In spite of
this, we notice that, the temperature dependent integrated intensity
of superlattice peak in TbTe$_3$ shows an evident dip near 150 K,
\cite{2CDW} which was considered as an experimental artifact by the
authors. However, it is very likely that the feature actually
signals the presence of the third CDW transition. In the early ARPES
measurement, the removal gaps on the FS sheets connected by the
nesting wave vector were claimed to be unequal, \cite{ARPESSmTe3}
which is in sharp contrast to the identical results in the later
ARPES probe. \cite{R-evolution3} The seemingly controversial results
are most likely caused by the bilayer splitting which was
investigated in different resolution conditions.

\emph{R}Te$_3$ systems represent a kind of rare compound which
experiences multiple CDW transitions. The other examples include
NbSe$_3$ and $\eta$-Mo$_4$O$_{11}$. Both exhibit two incommensurate
CDWs with \emph{$T_{c1}$} = 145 K and \emph{$T_{c2}$} = 59 K for
NbSe$_3$ \cite{NbSe31,NbSe32} and \emph{$T_{c1}$} = 109 K and
\emph{$T_{c2}$} = 30 K for $\eta$-Mo$_4$O$_{11}$.
\cite{chemical-review} Whereas, \emph{R}Te$_3$ is quite striking in
which as many as three distinct CDW orders emerge coincidentally. It
will be very interesting to study the interplay and relationship
among the three but the same type orders. In Fig. 3(a) we note that
gap 1 and gap 3 coexist in most compounds in the series and display
the similar monotonic evolution from light rare-earth to heavier
ones. Since the CDW orders are nesting driven, certain difference in
the nesting conditions for the two CDW orders should exist. It is
noted that the AB actually exhibits stronger
\emph{\textbf{b}$^{\ast}$}-axis dispersion (perpendicular to the
conduction \emph{ac}-plane) and thus worse nesting conditions than
BB. \cite{2CDW,R-evolution3} Therefore, BB bears much stronger CDW
instability than AB. The gap amplitude and the affected spectral
weight of the first CDW order on BB are both much larger than that
of the third one on AB.

The second CDW order, which occurs in the four heavy rare-earth
compounds, displays monotonic evolution opposite to the other two
ones. The transition temperature \emph{T$_{c2}$}, as well as the CDW
gap, increases with enhanced chemical pressure, which is rather
peculiar since pressure generally suppresses CDW transitions.
\cite{quenchCDW,CDW-pressure1} In Fig. 3(d), we notice that, from
LaTe$_3$ to TbTe$_3$, in which the second CDW order is absent, the
lost spectral weight \emph{W1} and \emph{W3} display rather little
variations. The rapid suppression of both values just right
coincides with the onset of the second CDW order. To examine the
relationship of the orders, we plot \emph{W1} versus \emph{W2} in
the inset. The two quantities show almost perfect linear
anti-correlation in the four heavy rare-earth compounds. It
explicitly demonstrates that the second CDW order competes with the
other two ones for the low energy spectral weight available for
nesting. With increased chemical pressure, the amount of gapped FS
by the first and third CDW orders reduces \cite{R-evolution3} and
more intact FS is left to contribute to the second CDW transition.
\cite{2CDW}

For the four heavy rare-earth compounds, the second CDW order
occurring perpendicular to the other two ones, a bidirectional
checkerboard CDW ground state would arise. \cite{2CDW,theoryRTe32}
It is reminiscent of the pseudogap state in high-\emph{T$_c$}
cuprates, of which the origin has been long debated between the
precursor superconducting paring gap and competing orders. In the
former viewpoint, the pseudogap is believed to be the preformed
Cooper pairs' gap before the coherence necessary for
superconductivity (SC), which is rigidly tied to the superconducting
phenomenon. In the latter point, the pseudogap is suggested to be a
new phase, having no direct relationship with SC, which even
competes and is harmful to SC. Though the origin is still under
debate, there is growing evidence that the pseudogap arises from a
checkerboard CDW order with perpendicular wave vectors close to
$\textbf{Q} = (\pm2\pi/4, 0), (0, \pm2\pi/4)$.
\cite{checkerboard1,checkerboard2,checkerboard3} The perpendicular
CDW orders have 4-folded symmetry \cite{checkerboard2,checkerboard3}
and occur simultaneously. While in the present systems, the 4-folded
symmetry was broken due to the weakly orthorhombic structure.
\cite{structure1,structure2} The distinction between the
perpendicular orders, nevertheless, tends to vanish with increased
(chemical) pressures. \cite{R-evolution2} A 4-folded checkerboard
CDW is expected in the tetragonal lattice under sufficient high
pressures.

\section{\label{sec:level2}CONCLUSIONS}
To conclude, we performed a systematic optical spectroscopy study
on CDWs in the eleven rare-earth tri-telluride compounds
\emph{R}Te$_3$ (\emph{R} = La - Nd, Sm, Gd - Tm). Besides the prior
reported two CDW orders, the study reveals unexpectedly the presence
of a third CDW order in the series which evolves systematically with
the size of \emph{R} element. The puzzling energy gap features
observed previously in the light rare-earth based compounds CeTe$_3$
and TbTe$_3$ at lower energies actually belong to this third CDW
order. With increased chemical pressure, the first and third CDW
orders are both substantially suppressed and compete with the second
one by depleting the low energy spectral weight. We suggest that the
third CDW order arises from the bilayer splitting, which lifts the
degeneracy of conduction bands of double Te sheets. The study
establishes a complete phase diagram for the multiple CDW orders in
this series.

\begin{center}
\small{\textbf{ACKNOWLEDGMENTS}}
\end{center}
This work was supported by the National Science Foundation of China
(11120101003, 11327806), and the 973 project of the Ministry of Science
and Technology of China (2011CB921701, 2012CB821403).

\end{document}